\documentclass[12pt]{article}
\usepackage[dvips]{graphicx}

\newcommand{\cd}{\makebox[0.08cm]{$\cdot$}}
\newcommand{\MSbar} {\hbox{$\overline{\hbox{\tiny MS}}$}}
\newcommand{\barMS} {\overline{\rm MS}}
\begin {document}
\begin{flushright}
{\small
SLAC--PUB--10576\\
July 2004\\}
\end{flushright}

\begin{center}
{{\bf\LARGE Conformal Symmetry as a Template for\\[1ex]
QCD}\footnote{Work supported by Department of Energy contract
DE--AC03--76SF00515.}}

\bigskip
{\it Stanley J. Brodsky \\
Stanford Linear Accelerator Center \\
Stanford University, Stanford, California 94309 \\
E-mail:  sjbth@slac.stanford.edu}
\medskip
\end{center}

\vfill

\begin{center}
{\bf\large Abstract }
\end{center}

Conformal symmetry is broken in physical QCD; nevertheless, one
can use conformal symmetry as a template, systematically
correcting for its nonzero $\beta$ function as well as
higher-twist effects.  For example, commensurate scale relations
which relate QCD observables to each other, such as the
generalized Crewther relation, have no renormalization scale or
scheme ambiguity and retain a convergent perturbative structure
which reflects the underlying conformal symmetry of the classical
theory. The ``conformal correspondence principle" also dictates
the form of the expansion basis for hadronic distribution
amplitudes.  The AdS/CFT correspondence connecting superstring
theory to superconformal gauge theory has important implications
for hadron phenomenology in the conformal limit, including an
all-orders demonstration of  counting rules for hard exclusive
processes as well as determining essential aspects of hadronic
light-front wavefunctions. Theoretical and phenomenological
evidence is now accumulating that QCD couplings based on physical
observables such as $\tau$ decay  become constant at small
virtuality; i.e., effective charges develop an infrared fixed
point in contradiction to the usual assumption of singular growth
in the infrared. The near-constant behavior of effective couplings
also suggests that QCD can be approximated as a conformal theory
even at relatively small momentum transfer. The importance of
using an analytic effective charge such as the pinch scheme for
unifying the electroweak and strong couplings and forces is also
emphasized.

\vfill

\begin{center}
{\it Invited talk given at the \\
Workshop Continuous Advances in QCD 2004\\
William I. Fine Theoretical Physics Institute\\
University of Minnesota, Minneapolis\\
May 13--16, 2004}
\end{center}

\vfill \newpage

\section{Introduction: The Conformal Correspondence Principle}

The classical Lagrangian of QCD for massless quarks is conformally symmetric.  Since it
has no intrinsic mass scale, the classical theory is invariant under the $SO(4,2)$
translations, boosts, and rotations of the Poincare  group, plus the dilatations and
other transformations of the conformal group. Scale invariance and therefore conformal
symmetry is destroyed in the quantum theory by the renormalization procedure which
introduces a renormalization scale as well as by quark masses. Conversely,
Parisi~\cite{Parisi:zy} has shown that perturbative QCD becomes a conformal theory  for
$\beta \to 0$ and zero quark mass. Conformal symmetry is thus broken in physical QCD;
nevertheless, we can still recover the underlying features of the conformally invariant
theory by evaluating any expression in QCD in the analytic limit of zero quark mass and
zero $\beta$ function:
\begin{equation}
\lim_{m_q \to 0, \beta \to 0} \mathcal{O}_{QCD} = \mathcal{ O}_{\rm conformal\ QCD} \ .
\end{equation} This conformal correspondence limit is analogous
to Bohr's correspondence principle where one recovers predictions of classical theory
from quantum theory in the limit of zero Planck constant.  The contributions to an
expression in QCD from its nonzero $\beta$-function can be systematically
identified~\cite{Brodsky:2000cr,Rathsman:2001xe,Grunberg:2001bz} order-by-order in
perturbation theory using the Banks-Zaks procedure~\cite{Banks:1981nn}.

There are a number of useful phenomenological consequences of near conformal behavior of
QCD: the conformal approximation with zero $\beta$ function can be used as template for
QCD analyses~\cite{Brodsky:1985ve,Brodsky:1984xk} such as the form of the expansion
polynomials for distribution amplitudes~\cite{Braun:2003rp,Braun:1999te}.  The
near-conformal behavior of QCD is the basis for commensurate scale
relations~\cite{Brodsky:1994eh} which relate observables to each other without
renormalization scale or scheme ambiguities~\cite{Brodsky:2000cr,Rathsman:2001xe}. By
definition, all contributions from the nonzero $\beta$ function can be incorporated into
the QCD running coupling $\alpha_s(Q)$ where $Q$ represents the set of physical
invariants.  Conformal symmetry thus provides a template for physical QCD expressions.
For example, perturbative expansions in QCD for massless quarks must have the form
\begin{equation}
 \mathcal{O} = \sum_{n=0} C_n \alpha^n_s(Q^*_n)
 \end{equation}
where the $C_n$ are identical to the expansion coefficients in the conformal theory, and
$Q^*_n$ is the scale chosen to resum all of the contributions from the nonzero $\beta$
function at that order in perturbation theory. Since the conformal theory does not
contain renormalons, the $C_n$ do not have the divergent $n!$ growth characteristic of
conventional PQCD expansions evaluated at a fixed scale.

\section{Effective Charges}

One can define the fundamental coupling of QCD from virtually any physical
observable~\cite{Grunberg:1980ja}. Such couplings, called ``effective charges", are
all-order resummations of perturbation theory, so they  correspond to the complete theory
of QCD. Unlike the $\barMS$ coupling, a physical coupling is analytic across quark flavor
thresholds~\cite{Brodsky:1998mf,Brodsky:1999fr}.  In particular, heavy particles will
contribute to physical predictions even at energies below their threshold.  This is in
contrast to mathematical renormalization schemes such as ${\bar MS},$ where mass
thresholds are treated as step functions. In addition, since the QCD running couplings
defined from observables are bounded, integrations over  effective charges are well
defined and the arguments requiring renormalon resummations do apply. The physical
couplings satisfy the standard renormalization group equation for its logarithmic
derivative, ${{\rm d}\alpha_{\rm phys}/{\rm d}\ln k^2} = \widehat{\beta}_{\rm
phys}[\alpha_{\rm phys}(k^2)]$, where the first two terms in the perturbative expansion
of $\widehat{\beta}_{\rm phys}$ are scheme-independent at leading twist; the higher order
terms have to be calculated for each observable separately using perturbation theory.

Commensurate scale relations are QCD predictions which relate observables to each other
at their respective scales. An important example is the generalized Crewther
relation~\cite{Brodsky:1995tb}:
\begin{equation}
\left[1 + \frac{\alpha_R(s^*)}{\pi} \right] \left[1 -
\frac{\alpha_{g_1}(Q^2)}{\pi}\right] = 1
\end{equation}
where the underlying form at zero $\beta$ function is dictated by conformal
symmetry~\cite{Crewther:1972kn}. Here $\alpha_R(s)/\pi$ and $-\alpha_{g_1}(Q^2)/\pi$
represent the entire radiative corrections to $R_{e^+ e^-}(s)$ and the Bjorken sum rule
for the $g_1(x,Q^2)$ structure function measured in spin-dependent deep inelastic
scattering, respectively. The relation between $s^*$ and $Q^2$ can be computed order by
order in perturbation theory, as in the BLM method~\cite{Brodsky:1982gc}.   The ratio of
physical scales guarantees that the effect of new quark thresholds is commensurate.
Commensurate scale relations are renormalization-scheme independent and satisfy the group
properties of the renormalization group.  Each observable can be computed in any
convenient renormalization scheme such as dimensional regularization. The $\barMS$
coupling can then be eliminated; it becomes only an intermediary~\cite{Brodsky:1994eh}.
In such a procedure there are no further renormalization scale ($\mu$) or scheme
ambiguities.

In the case of QED,  the heavy lepton potential (in the limit of vanishing external
charge) is conventionally used to define the effective charge $\alpha_{qed}(q^2)$.  This
definition, the Dyson Goldberger-Low effective charge, resums all lepton pair vacuum
polarization contributions in the photon propagator, and it is analytic in the lepton
masses.  The scale of the QCD coupling is thus the virtuality of the exchanged photon.
The extension of this concept to non-abelian gauge theories is non-trivial due to the
self interactions of the gauge bosons which make the usual self-energy gauge dependent.
However, by systematically implementing the Ward identities of the theory, one can
project out the unique self-energy of each {\it physical} particle.  This results in a
gluonic self-energy which is gauge independent and which can be resummed to define an
effective charge that is related through the optical theorem to differential cross
sections. The algorithm for performing the calculation at the diagrammatic level is
called the ``pinch technique"~\cite{Cornwall:1981zr,Degrassi:1992ue,watson,prw}. The
generalization of the pinch technique to higher loops has recently been
investigated~\cite{Watson:1998vw,watson2b,rafael2,Binosi:2002ft,Binosiqcdall,Binosi:2004qe}.
Binosi and Papavassiliou~\cite{Binosi:2002ft,Binosiqcdall,Binosi:2004qe} have shown the
consistency of the pinch technique to all orders in perturbation theory, thus allowing a
systematic application to the QCD and electroweak effective charges at higher orders. The
pinch scheme is in fact used to define the evolution of the couplings in the electroweak
theory.  The pinch scheme thus provides an ideal scheme for QCD couplings as well.

\section{Effective Charges and Unification}

\begin{figure}[htb]
 \centering
 \includegraphics[height=3in]{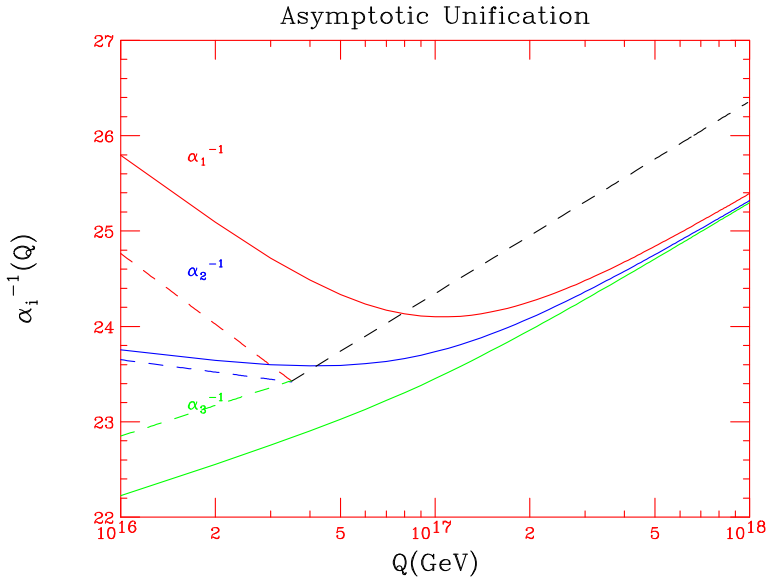}
 \caption[*]{
{\bf Asymptotic Unification.}  An illustration of strong and electroweak coupling
unification in an $SU(5)$ supersymmetric model based on the pinch scheme effective
charge.  The solid lines are the analytic pinch scheme $\overline{PT}$ effective
couplings, while the dashed lines are the $\overline{DR}$ couplings.  For illustrative
purposes, $a_3(M_Z)$ has been chosen  so that unification occurs at a finite scale for
$\overline{DR}$ and asymptotically for the $\overline{PT}$ couplings.  Here
$M_{SUSY}=200{\rm GeV}$ is the mass of all light superpartners except the wino and gluino
which have values $ \frac{1}{2}{mgx} = M_{SUSY} = 2 m_{wx}$.} \label{fig:4}
\end{figure}

Recently Michael Binger and I have analyzed a supersymmetric grand unification model in
the context of physical renormalization schemes~\cite{Binger:2003by}.  Our essential
assumption is that the underlying forces of the theory become at the unification scale.
We have found a number of qualitative differences and improvements in precision over
conventional approaches.  There is no need to assume that the particle spectrum has any
specific structure; the effect of heavy particles is included both below and below the
physical threshold.  Unlike mathematical schemes such as dimensional reduction,
$\overline{DR}$, the evolution of the coupling is analytic and  unification is approached
continuously rather than at a fixed scale. The effective charge formalism thus provides a
template for calculating all mass threshold effects for any given grand unified theory.
These new threshold corrections  are important in making the measured values of the gauge
couplings consistent with unification. A comparison with the conventional scheme based on
$\overline{DR}$ dimensional regularization scheme is summarized in Fig.~\ref{fig:4}.

\section{The Infrared Behavior of Effective QCD Couplings}

It is often assumed that color confinement in QCD can be traced to the singular behavior
of the running coupling in the infrared, {\em i.e.} ``infrared slavery."  For example if
$\alpha_s(q^2) \to {1}/{q^2}$ at $q^2 \to 0$, then one-gluon exchange leads to a linear
potential at large distances.  However,
theoretical~\cite{vonSmekal:1997is,Zwanziger:2003cf,Howe:2002rb,Howe:2003mp,%
Furui:2003mz} and
phenomenological~\cite{Mattingly:ej,Brodsky:2002nb,Baldicchi:2002qm}
evidence is now accumulating that the QCD coupling becomes
constant at small virtuality; {\em i.e.}, $\alpha_s(Q^2)$ develops
an infrared fixed point in contradiction to the usual assumption
of singular growth in the infrared.  Since all observables are
related by commensurate scale relations, they all should have an
IR fixed point~\cite{Howe:2003mp}.  A recent study of the QCD
coupling using lattice gauge theory in Landau gauge in fact shows
an infrared fixed point~\cite{Furui:2004bq}.  This result is also
consistent with Dyson-Schwinger equation studies of the physical
gluon propagator~\cite{vonSmekal:1997is,Zwanziger:2003cf}.  The
relationship of these results to the infrared-finite coupling for
the vector interaction defined in the quarkonium potential has
recently been discussed by Badalian and
Veselov~\cite{Badalian:2004ig}.

Menke, Merino, and Rathsman~\cite{Brodsky:2002nb} and I have
considered a  physical coupling for QCD which is defined
from the high precision measurements of the hadronic
decay channels of the $\tau^- \to \nu_\tau {\rm h}^-$.  Let
$R_{\tau}$ be the ratio of the hadronic decay rate to the leptonic
rate.  Then $R_{\tau}\equiv
R_{\tau}^0\left[1+\frac{\alpha_\tau}{\pi}\right]$, where
$R_{\tau}^0$ is the zeroth order QCD prediction, defines the
effective charge $\alpha_\tau$.  The data for $\tau$ decays is
well-understood channel by channel, thus allowing the calculation
of the hadronic decay rate and the effective charge as a function
of the $\tau$ mass below the physical mass.  The vector and
axial-vector decay modes which can be studied separately.
Using an analysis of the $\tau$ data from the OPAL
collaboration~\cite{Ackerstaff:1998yj}, we have found that the
experimental value of the coupling $\alpha_{\tau}(s)=0.621 \pm
0.008$ at $s = m^2_\tau$ corresponds to a value of
$\alpha_{\MSbar}(M^2_Z) = (0.117$-$0.122) \pm 0.002$, where the
range corresponds to three different perturbative methods used in
analyzing the data.  This result is in good agreement with the
world average $\alpha_{\MSbar}(M^2_Z) = 0.117 \pm 0.002$.  However,
from the figure we also see that the effective charge only reaches
$\alpha_{\tau}(s) \sim 0.9 \pm 0.1$ at $s=1\,{\rm GeV}^2$, and it
even stays within the same range down to $s\sim0.5\,{\rm GeV}^2$.
This result is in good agreement with the estimate of Mattingly
and Stevenson~\cite{Mattingly:ej} for the effective coupling
$\alpha_R(s) \sim 0.85 $ for $\sqrt s < 0.3\,{\rm GeV}$ determined
from ${\rm e}^+{\rm e}^-$ annihilation, especially if one takes
into account the perturbative commensurate scale relation,
$\alpha_{\tau}(m_{\tau^\prime}^2)= \alpha_R(s^*),$ where $s^*
\simeq 0.10\,m_{\tau^\prime}^2.$ This behavior is not consistent
with the coupling having a Landau pole, but rather shows that the
physical coupling is close to constant at low scales, suggesting
that physical QCD couplings are effectively constant or ``frozen"
at low scales.

\begin{figure}[htb]
\centering
\includegraphics[width=4.3in]   
{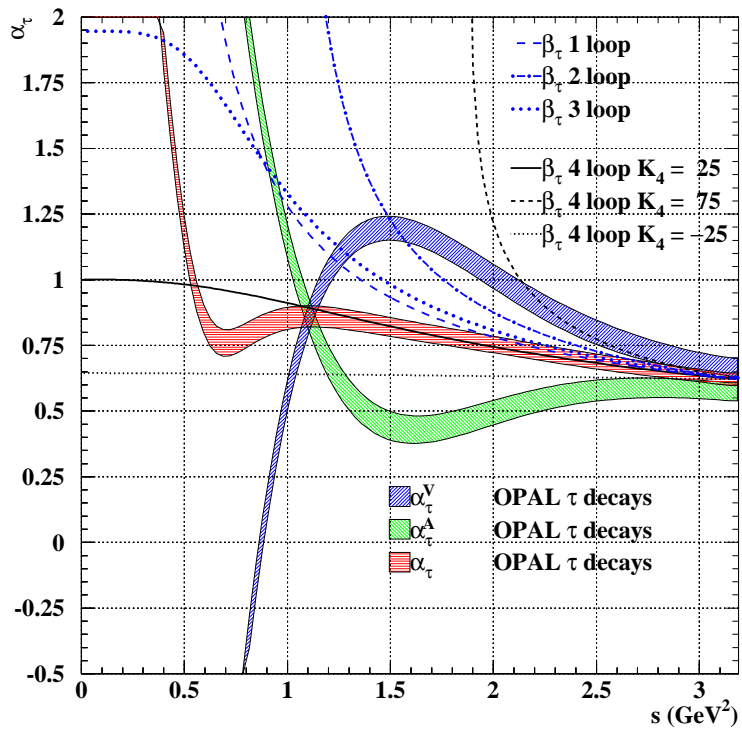} \caption[*]{The effective charge $\alpha_{\tau}$ for
non-strange hadronic decays of a hypothetical $\mathit{\tau}$
lepton with $\mathit{m_{\tau'}^2 = s}$ compared to solutions of
the fixed order evolution equation for $\alpha_{\tau}$ at two-,
three-, and four-loop order.  The error bands include statistical
and systematic errors. \label{fig:fopt_comp}}
\end{figure}

Figure~\ref{fig:fopt_comp} shows a comparison of the
experimentally determined effective charge $\alpha_{\tau}(s)$ with
solutions to the evolution equation for $\alpha_{\tau}$ at two-,
\hbox{three-,} and four-loop order normalized at $m_\tau$.  At
three loops the behavior of the perturbative solution drastically
changes, and instead of diverging, it freezes to a value
$\alpha_{\tau}\simeq 2$ in the infrared. The infrared behavior is
not perturbatively stable since the evolution of the coupling is
governed by the highest order term.  This is illustrated by the
widely different results obtained for three different values of
the unknown four loop term $\beta_{\tau,3}$ which are also shown.
The values of $\beta_{\tau,3}$ used are obtained from the estimate
of the four loop term in the perturbative series of $R_\tau$,
$K_4^{\overline{\rm MS}} = 25\pm 50$~\cite{LeDiberder:1992fr}. It
is interesting to note that the central four-loop solution is in
good agreement with the data all the way down to $s\simeq1\,{\rm
GeV}^2$.

The results for $\alpha_{\tau}$ resemble the behavior of the
one-loop ``time-like" effective
coupling~\cite{Beneke:1994qe,Ball:1995ni,Dokshitzer:1995qm}
\begin{equation}\label{eq:alphaeff}
\alpha_{\rm eff}(s)=\frac{4\pi}{\beta_0} \left\{\frac{1}{2} -
\frac{1}{\pi}\arctan\left[\frac{1}{\pi}\ln\frac{s}{\Lambda^2}\right]\right\}
\end{equation}
which is finite in the infrared and freezes to the value
$\alpha_{\rm eff}(s)={4\pi}/{\beta_0}$ as $s\to 0$.  It is
instructive to expand the ``time-like" effective coupling for
large $s$,
\begin{eqnarray}
\alpha_{\rm eff}(s)
&=&\frac{4\pi}{\beta_0\ln\left(s/\Lambda^2\right)} \left\{1
-\frac{1}{3}\frac{\pi^2}{\ln^2\left(s/\Lambda^2\right)}
+\frac{1}{5}\frac{\pi^4}{\ln^4\left(s/\Lambda^2\right)} +\ldots
\right\} \nonumber\\ &=&\alpha_{\rm s}(s)\left\{1
-\frac{\pi^2\beta_0^2}{3}\left(\frac{\alpha_{\rm
s}(s)}{4\pi}\right)^2
+\frac{\pi^4\beta_0^4}{5}\left(\frac{\alpha_{\rm
s}(s)}{4\pi}\right)^4 +\ldots \right\}.
\end{eqnarray}
This shows that the ``time-like" effective coupling is a
resummation of $(\pi^2\beta_0^2\alpha_{\rm s}^2)^n$-corrections to
the usual running couplings.  The finite coupling $\alpha_{\rm
eff}$ given in Eq.~(\ref{eq:alphaeff}) obeys standard PQCD
evolution at LO.  Thus one can have a solution for the
perturbative running of the QCD coupling which obeys asymptotic
freedom but does not have a Landau singularity.

The near constancy of the effective QCD coupling at small scales
illustrates the near-conformal behavior of QCD.  It
helps explain the empirical success of dimensional counting rules
for the power law fall-off of form factors and fixed angle
scaling.  As shown in the
references~\cite{Brodsky:1997dh,Melic:2001wb}, one can calculate
the hard scattering amplitude $T_H$ for such
processes~\cite{Lepage:1980fj} without scale ambiguity in terms of
the effective charge $\alpha_\tau$ or $\alpha_R$ using
commensurate scale relations.  The effective coupling is evaluated
in the regime where the coupling is approximately constant, in
contrast to the rapidly varying behavior from powers of
$\alpha_{\rm s}$ predicted by perturbation theory (the universal
two-loop coupling).  For example, the nucleon form factors are
proportional at leading order to two powers of $\alpha_{\rm s}$
evaluated at low scales in addition to two powers of $1/q^2$; The
pion photoproduction amplitude at fixed angles is proportional at
leading order to three powers of the QCD coupling.  The essential
variation from leading-twist counting-rule behavior then only
arises from the anomalous dimensions of the hadron distribution
amplitudes.

\section{Light-Front Quantization}

The concept of a wave function of a hadron as a composite of
relativistic quarks and gluons is naturally formulated in terms of
the light-front Fock expansion at fixed light-front time, $\tau=x
\cd \omega$.  The four-vector $\omega$, with $\omega^2 = 0$,
determines the orientation of the light-front plane; the freedom
to choose $\omega$ provides an explicitly covariant formulation of
light-front quantization~\cite{cdkm}.  Although LFWFs depend on
the choice of the light-front quantization direction, all
observables such as matrix elements of local current operators,
form factors, and cross sections are light-front invariants --
they must be independent of $\omega_\mu.$

The light-front wave functions (LFWFs) $\psi_n(x_i,k_{\perp_i},\lambda_i)$, with
$x_i=\frac{k_i \cd \omega}{P\cd \omega}$, $\sum^n_{i=1} x_i=1, $
$\sum^n_{i=1}k_{\perp_i}=0_\perp$, are the coefficient functions for $n$ partons in the
Fock expansion, providing a general frame-independent representation of the hadron state.
Matrix elements of local operators such as spacelike proton form factors can be computed
simply from the overlap integrals of light front wave functions in analogy to
nonrelativistic Schr\"odinger theory. In principle, one can solve for the LFWFs directly
from fundamental theory using nonperturbative methods such as discretized light-front
quantization (DLCQ), the transverse lattice, lattice gauge theory moments, or
Bethe--Salpeter techniques. The determination of the hadron LFWFs from phenomenological
constraints and from QCD itself is a central goal of hadron and nuclear physics.  Reviews
of nonperturbative light-front methods may be found in the
references~\cite{Brodsky:1997de,cdkm,Dalley:ug,Brodsky:2003gk}. A potentially important
method is to construct the $q\bar q$ Green's function using light-front Hamiltonian
theory, with DLCQ boundary conditions and Lippmann-Schwinger resummation.  The zeros of
the resulting resolvent projected on states of specific angular momentum $J_z$ can then
generate the meson spectrum and their light-front Fock wavefunctions.  The DLCQ
properties and boundary conditions allow a truncation of the Fock space while retaining
the kinematic boost and Lorentz invariance of light-front quantization.

One of the central issues in the analysis of fundamental hadron
structure is the presence of non-zero orbital angular momentum in
the bound-state wave functions.  The evidence for a ``spin crisis"
in the Ellis-Jaffe sum rule signals a significant orbital
contribution in the proton wave
function~\cite{Jaffe:1989jz,Ji:2002qa}.  The Pauli form factor of
nucleons is computed from the overlap of LFWFs differing by one
unit of orbital angular momentum $\Delta L_z= \pm 1$.  Thus the
fact that the anomalous moment of the proton is non-zero requires
nonzero orbital angular momentum in the proton
wavefunction~\cite{BD80}.  In the light-front method, orbital
angular momentum is treated explicitly; it includes the orbital
contributions induced by relativistic effects, such as the
spin-orbit effects normally associated with the conventional Dirac
spinors.

In recent work, Dae Sung Hwang, John Hiller, Volodya Karmonov~\cite{Brodsky:2003pw}, and
I  have studied the analytic structure of LFWFs using the explicitly Lorentz-invariant
formulation of the front form.  Eigensolutions of the Bethe-Salpeter equation have
specific angular momentum as specified by the Pauli-Lubanski vector.  The corresponding
LFWF for an $n$-particle Fock state evaluated at equal light-front time $\tau =
\omega\cdot x$ can be obtained by integrating the Bethe-Salpeter solutions over the
corresponding relative light-front energies.  The resulting LFWFs $\psi^I_n(x_i, k_{\perp
i})$ are functions of the light-cone momentum fractions $x_i= {k_i\cdot \omega / p \cdot
\omega}$ and the invariant mass squared of the constituents $M_0^2= (\sum^n_{i=1}
k_i^\mu)^2 =\sum_{i =1}^n \big [\frac{k^2_\perp + m^2}{x}\big]_i$ and the light-cone
momentum fractions $x_i= {k\cdot \omega / p \cdot \omega}$ each multiplying spin-vector
and polarization tensor invariants which can involve $\omega^\mu.$  The resulting LFWFs
for bound states are eigenstates of the Karmanov--Smirnov kinematic angular momentum
operator~\cite{ks92} and satisfy all of the Lorentz symmetries of the front form,
including boost invariance.

\section{AFS/CFT Correspondence and Hadronic Light-Front Wavefunctions}

As shown by Maldacena~\cite{Maldacena:1997re}, there is a remarkable correspondence
between large $N_C$ supergravity theory in a higher dimensional  anti-de Sitter space and
supersymmetric QCD in 4-dimensional space-time.  String/gauge duality provides a
framework for predicting QCD phenomena based on the conformal properties of the AdS/CFT
correspondence. The AdS/CFT correspondence is based on the fact that the generators of
conformal and Poincare transformations have representations on the five-dimensional
anti-deSitter space $AdS_5$  as well as Minkowski spacetime. For example, Polchinski and
Strassler~\cite{Polchinski:2001tt} have shown that the power-law fall-off of hard
exclusive hadron-hadron scattering amplitudes at large momentum transfer can be derived
without the use of perturbation theory by using the scaling properties of the hadronic
interpolating fields in the large-$r$ region of  AdS space.  Thus one can use the
Maldacena correspondence to compute the leading power-law falloff of exclusive processes
such as high-energy fixed-angle scattering of gluonium-gluonium scattering in
supersymmetric QCD.   The resulting predictions for hadron physics effectively
coincide~\cite{Polchinski:2001tt,Brower:2002er,Andreev:2002aw} with QCD dimensional
counting rules:\cite{Brodsky:1973kr,Matveev:ra,Brodsky:1974vy}
\begin{equation}
\frac{d\sigma}{dt}(H_1 H_2 \to H_3 H_4) =\frac{ F(t/s)}{s^{n-2}}\end{equation} where $n$
is the sum of the minimal number of interpolating fields in the initial and final state.
(For a recent review of hard fixed $\theta_{CM}$ angle exclusive processes in QCD see
reference~\cite{Brodsky:2002st}.) As shown by Brower and Tan~\cite{Brower:2002er}, the
non-conformal dimensional scale which appears in the QCD analysis is set by the string
constant, the  slope of the primary Regge trajectory $\Lambda^2=\alpha^\prime_R(0)$ of
the supergravity theory. Polchinski and Strassler~\cite{Polchinski:2001tt} have also
derived counting rules for deep inelastic structure functions at $x \to 1$ in agreement
with perturbative QCD predictions~\cite{Brodsky:1994kg} as well as Bloom-Gilman
exclusive-inclusive duality.

The supergravity analysis is based on an extension of classical gravity theory in higher
dimensions and is nonperturbative.  Thus analyses of exclusive
processes~\cite{Lepage:1980fj} which were based on perturbation theory can be extended by
the Maldacena correspondence to all orders.  An interesting point is that the hard
scattering amplitudes which are normally or order $\alpha_s^p$ in PQCD appear as order
$\alpha_s^{p/2}$ in the supergravity predictions.  This can be understood as an
all-orders resummation of the effective potential~\cite{Maldacena:1997re,Rey:1998ik}.

The superstring theory results are derived in the limit of a large
$N_C$~\cite{'tHooft:1973jz}.  For gluon-gluon scattering, the amplitude scales as
${1}/{N_C^2}$.   For color-singlet bound states of quarks, the amplitude scales as
${1}/{N_C}$.  This large $N_C$-counting in fact corresponds to the quark interchange
mechanism~\cite{Gunion:1973ex}.  For example, for $K^+ p \to K^+ p$ scattering, the
$u$-quark exchange amplitude scales approximately as $\frac{1}{u}$\ $\frac{1}{t^2},$
which agrees remarkably well with the measured large $\theta_{CM}$ dependence of the $K^+
p$ differential cross section~\cite{Sivers:1975dg}. This implies that the nonsinglet
Reggeon trajectory asymptotes to a negative integer~\cite{Blankenbecler:1973kt}, in this
case, $\lim_{-t \to \infty}\alpha_R(t) \to -1.$

De Teramond and I~\cite{Brodsky:2003px} have shown how to compute the form and scaling of
light-front hadronic wavefunctions using the AdS/CFT correspondence in quantum field
theories which  have an underlying conformal structure, such as ${\mathcal N} = 4$
super-conformal QCD.  For example, baryons are included in the theory by adding an open
string sector in $AdS_5 \times S^5$ corresponding to quarks in the fundamental
representation of  the $SU(4)$ symmetry defined on $S^5$ and the fundamental and higher
representations of $SU(N_C).$ The hadron mass scale is introduced by imposing boundary
conditions at the $AdS_5$ coordinate  $r= r_0 = \Lambda_{QCD} R^2.$ The quantum numbers
of the lowest Fock state of each hadron including its internal orbital angular momentum
and spin-flavor symmetry, are identified by matching the fall-off of the string
wavefunction  $\Psi(x,r)$ at the asymptotic $3+1$ boundary. Higher Fock states are
identified with conformally invariant quantum fluctuations of the bulk geometry about the
AdS background. The scaling and conformal properties of the AdS/CFT correspondence leads
to a hard component of the LFWFs of the form:
\begin{eqnarray}
\psi_{n/h} (x_i, \vec k_{\perp i} , \lambda_i, l_{z i}) &\sim&
\frac{(g_s~N_C)^{\frac{1}{2} (n-1)}}{\sqrt {N_C}} ~\prod_{i =1}^{n
- 1} (k_{i \perp}^\pm)^{\vert l_{z i}\vert} ~ \nonumber \\[1ex]
&\times&\left[\frac{ \Lambda_o}{ {M}^2 - \sum _i\frac{\vec k_{\perp i}^2 + m_i^2}{x_i} +
\Lambda_o^2}  \right] ^{n +\vert l_z \vert -1}, \label{eq:lfwfR}
\end{eqnarray}
where $g_s$ is the string scale and $\Lambda_o$ represents the basic QCD mass scale.  The
scaling predictions agree with  perturbative QCD analyses~\cite{Ji:bw,Lepage:1980fj}, but
the AdS/CFT analysis is performed at strong coupling without the use of perturbation
theory. The near-conformal scaling properties of light-front wavefunctions lead to a
number of other predictions for QCD which are normally discussed in the context of
perturbation theory, such as constituent counting scaling laws for the leading power
fall-off of form factors and hard exclusive scattering amplitudes for QCD processes.  The
ratio of Pauli to Dirac baryon form factor have the nominal asymptotic form ${F_2(Q^2) /
F_1(Q^2) }\sim 1/Q^2$, modulo logarithmic corrections, in agreement with the perturbative
results~\cite{Belitsky:2002kj}.  Our analysis can also be extended to study the spin
structure of scattering amplitudes at large transverse momentum and other processes which
are dependent on the scaling and orbital angular momentum structure of light-front
wavefunctions.

\section*{Acknowledgements}

This talk is based on collaborations with Michael Binger, Gregory Gabadadze, Einan Gardi,
Georges Grunberg, John Hiller, Dae Sung Hwang, Volodya Karmanov, Andre Kataev, Hung Jung
Lu, Sven Menke, Carlos Merino, Johan Rathsman, and Guy de T\'eramond. I am also thankful
to the William I. Fine Theoretical Physics Institute at the University of Minnesota which
sponsored this meeting. This work was supported by the U.S. Department of Energy,
contract DE--AC03--76SF00515.

\end {document}